\documentclass[journal]{IEEEtran}
\usepackage{amsmath}
\usepackage{amsfonts}
\usepackage{float}
\usepackage{amssymb}
\usepackage[T1]{fontenc}
\usepackage{amsthm}
\usepackage{units}

\usepackage{cite}

\ifCLASSINFOpdf
  \usepackage[pdftex]{graphicx}
   \graphicspath{{../pdf/}{../jpeg/}}
  \DeclareGraphicsExtensions{.pdf,.jpeg,.png}
\else
  \usepackage[dvips]{graphicx}
  \graphicspath{{../eps/}}
  \DeclareGraphicsExtensions{.eps}
\fi
\ifCLASSINFOpdf
\else
\fi
\begin{document}
%
\title{Limited Feedback-Based Interference Alignment for Interfering Multi-Access
Channels}
%
%
%

\author{Hui~Gao,~\IEEEmembership{Member,~IEEE,}
        Tiejun~Lv,~\IEEEmembership{Senior Member,~IEEE,}
        Di~Fang,
        Shaoshi~Yang,~\IEEEmembership{Member,~IEEE,}
        and Chau~Yuen,~\IEEEmembership{Senior Member,~IEEE.}

\thanks{H. Gao, T. Lv and D. Fang are with the Key Laboratory of Trustworthy
Distributed Computing and Service, Ministry of Education, and the
School of Information and Communication Engineering, Beijing University
of Posts and Telecommunications, Beijing, China 100876 (e-mail: \{huigao, lvtiejun, fangdi\}@bupt.edu.cn). H. Gao is also with Singapore University of Technology and Design, 20 Dover Drive, Singapore 138682.}
\thanks{S. Yang is with the School of Electronics and Computer Science, University
of Southampton, SO17 1BJ Southampton, U.K. (e-mail: shaoshi.yang@soton.ac.uk).}
\thanks{C. Yuen is with Singapore University of Technology and Design, 20
Dover Drive, Singapore 138682 (email: yuenchau@sutd.edu.sg).}
\thanks{This work is partially supported by the National Natural Science
Foundation of China (NSFC) (Grant No. 61271188) and Singapore University of Technology and Design.}
}

%
%

\markboth{IEEE COMMUNICATIONS LETTERS, ACCEPTED FOR PUBLICATION}%
{Shell \MakeLowercase{\textit{et al.}}: Limited Feedback Based Interference Alignment for Interfering Multi-Access
Channels}
%



\maketitle

\begin{abstract}
A limited feedback-based interference alignment (IA) scheme is proposed for the interfering multi-access channel (IMAC). By employing a novel performance-oriented quantization strategy, the proposed scheme is able to achieve the minimum overall residual inter-cell interference (ICI) with the optimized transceivers under limited feedback. Consequently, the scheme outperforms the existing counterparts in terms of system throughput. In addition, the proposed scheme can be implemented with flexible antenna configurations.
\end{abstract}


%
\IEEEpeerreviewmaketitle

\section{Introduction}

Interference alignment (IA) \cite{cadambe2008interference,maddah-ali_communication_2008}
has been considered as a new paradigm of intelligent interference
management in wireless networks. The principle of IA is to align the
interference signals into a subspace with minimum dimensions at the
receiver, and consequently the degrees-of-freedoms (DoFs) of the desired
signals can be maximized. Recently, various IA schemes have been developed
for the interfering broadcast channel (IBC) and interfering multi-access
channel (IMAC) \cite{suh_interference_2008,Downlink_IA_2011,two-cell_mac_2011,tranceiver_2012},
which are typical models of the cellular networks. It is shown that
these IA schemes are capable of improving system throughput in the
presence of interference. However, the promise of IA is primarily
based on the assumption of global channel state information (CSI)
at all transceivers, which requires significant system overhead for
CSI.

Aiming at more practical implementations, many IA schemes with
limited feedback have been proposed \cite{cho_interference_2012,limited_twocell_mac_2012,Xiaoming1,JL,Huiwnnc1}.
Of particular interests are the IA schemes for the IMAC with limited
feedback \cite{cho_interference_2012,limited_twocell_mac_2012}, where
the base station (BS) generates the IA-inspired transmit beamforming (TB) vectors
with perfect CSI and then feeds back their quantizations to users for uplink transmission. Although
\cite{cho_interference_2012,limited_twocell_mac_2012} assume ideal
IA with perfect CSI, the actual inter-cell interference (ICI) cannot
be fully canceled at the BS because of the quantization errors. Therefore,
the residual ICI is inevitable with limited feedback, which dramatically
reduces the achievable system throughput. Moreover, the quantization
strategies in \cite{cho_interference_2012,limited_twocell_mac_2012}
are not able to fully exploit the potential of limited feedback. Although
the selected codewords are the best approximations of the IA-inspired
TB vectors within the given codebooks in terms of chordal
distance, such codewords are not necessarily the ones that minimize
the residual ICI, which directly influences the system performance. Finally, the transceiver designs in \cite{cho_interference_2012,limited_twocell_mac_2012}
are constrained by the feasibility of IA and are only applicable to some specific antenna configurations.

In this paper, we propose a limited feedback-based IA scheme for the two-cell
IMAC. In contrast to \cite{cho_interference_2012,limited_twocell_mac_2012},
the proposed scheme is capable of minimizing the overall residual
ICI of each cell with the given codebooks, and it can be implemented
with flexible antenna configuration. To be specific, in each cell the optimal
receive filter of BS is first derived with an arbitrary set of TB vectors of the users in the neighbor cell,
and then the overall residual ICI of each cell is transformed into
a single-variable function of the set of TB
vectors. Next, each BS jointly quantizes the TB vectors
with a compound codebook and a new criterion aiming to directly minimize the overall residual ICI. In this way, the optimized transceivers are obtained under the framework of limited feedback, which effectively approach perfect IA in terms of the minimum overall residual ICI with the given codebooks. Benefiting from the new quantization strategy, the achievable system throughput of our scheme is
significantly improved as compared with \cite{cho_interference_2012,limited_twocell_mac_2012}.
Finally, it is worth pointing out that the opportunistic IA in \cite{JL,Huiwnnc1}
are practical IA schemes with limited feedback, which exploit the
multi-user diversity inherited in the cellular networks for throughput gains; the joint design of the proposed scheme with opportunistic
user scheduling may be an interesting future direction%
\footnote{\textbf{Notation: }Bold upper case and lower case letters represent
to matrices and vectors, respectively. $\left(\,\cdot\,\right)^{H}$
denotes the Hermitian transpose and $\left(\,\cdot\,\right)^{-1}$
represents the inverse of a matrix. $\left\Vert \,\cdot\,\right\Vert $
represent the Frobenius norm of a matrix. $\mathbb{E}\left[\,\cdot\,\right]$
stands for the expectation. $\rho_{K}(\mathbf{A})$ and ${\rm {Col}}\left({\bf {A}}\right)$
denote the sum of the minimum $K$ eigenvalues of $\mathbf{A}$ and
the column space of $\mathbf{A}$.%
}.

\section{System Model}

\begin{figure}
\centering \includegraphics[height=3.5cm]{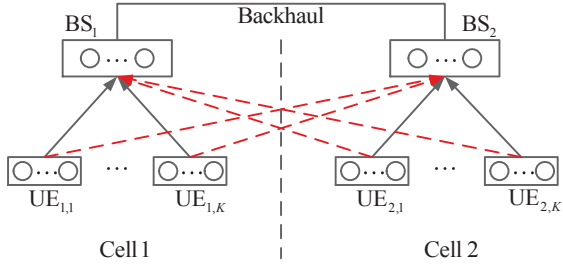} \caption{Two-cell IMAC with limited feeback via backhal. The solid and dashed lines represent
the desired and the interfering signals, respectively.}
\end{figure}

As illustrated in Fig. 1, a two-cell IMAC is considered, where
$\mathrm{B}\mathrm{S}_{i}$, $i\in\left\{ 1,2\right\} $, serves $K$
users $\left\{ \mathrm{UE}_{i,k}\right\} _{k=1}^{K}$ in the $i$-th
cell. It is assumed that the BS and the user are equipped with $N_{r}$
$(K<N_{r}<2K)$ and $N_{t}$ antennas, and each user transmits a single
data-stream. The direct and interference channels from
${{\mathop{\rm UE}\nolimits}_{i,k}}$ to its home ${{\mathop{\rm BS}\nolimits}_{i}}$
and its neighbor $\mathrm{BS}_{i'}$ are denoted as $\mathbf{H}_{i,k}\in\mathbb{C}^{N_{r}\times N_{t}}$
and $\mathbf{G}_{i,k}\in\mathbb{C}^{N_{r}\times N_{t}}$, respectively,
$i,\, i'\in\{1,2\}$, $i'\neq i$. The entries of $\mathbf{H}_{i,k}$
and $\mathbf{G}_{i,k}$ are assumed to be independently and identically
distributed (i.i.d.) complex Gaussian random variables with zero mean
and unit variance, i.e., $\mathcal{CN}(0,1)$. Frequency-division duplexing mode is assumed
in this system, and each user sends pilot symbols to both BSs for
CSI acquisition. Then $\mathrm{BS}_{i}$ selects the TB
vectors from the known codebooks for the users in the neighbor cell
by using the estimated CSI of interference channels. The corresponding
indices of codewords are exchanged between the two BSs via a backhaul
link and then fed back to users to instruct the selection of TB vectors.

During the uplink transmission, all users send information and the
received signal at $\mathrm{B}\mathrm{S}_{i}$ is given by
\begin{equation}
\mathbf{y}_{i}=\sum_{k=1}^{K}\mathbf{H}_{i,k}\mathbf{v}_{i,k}s_{i,k}+\sum_{n=1}^{K}\mathbf{G}_{i',n}\mathbf{v}_{i',n}s_{i',n}+\mathbf{n}_{i},
\end{equation}
where $\mathbf{v}_{i,k}\in\mathbb{C}^{N_{t}\times1}$ and $s_{i,k}\in\mathbb{C}$
respectively denote the normalized TB vector chosen from $\mathrm{UE}_{i,k}$'s codebook and the transmitted symbol of $\mathrm{UE}_{i,k}$
with an average power of $P$; $\mathbf{n}_{i}\in\mathbb{C}^{N_{r}\times1}$
is the noise vector with i.i.d. elements following $\mathcal{CN}(0,N_{0})$.
The receiver at $\mathrm{B}\mathrm{S}_{i}$ is a cascaded filter $\mathbf{r}_{i,k}^{H}\mathbf{U}_{i}^{H}$,
where $\mathbf{U}_{i}\in\mathbb{C}^{N_{r}\times K}$ and $\mathbf{r}_{i,k}\in\mathbb{C}^{K\times1}$
are designed to cancel the ICI and the intra-cell interference for
$\mathrm{UE}_{i,k}$. The column vectors of $\mathbf{U}_{i}$ are
normalized and orthogonal to each other. Let us define the effective channel
matrix from all the users $\left\{ \mathrm{UE}_{i,k}\right\} _{k=1}^{K}$
to $\mathrm{BS}_{i}$ as $\mathbf{H}_{i,e}=\mathbf{U}_{i}^{H}[\mathbf{H}_{i,1}\mathbf{v}_{i,1},...,\mathbf{H}_{i,K}\mathbf{v}_{i,K}]\in\mathbb{C}^{K\times K}$, $\mathbf{r}_{i,k}^{H}$ is then given by the $k$-th normalized row
vector of $\mathbf{H}_{i,e}^{-1}$. Based on (1) and the cascaded filter $\mathbf{r}_{i,k}^{H}\mathbf{U}_{i}^{H}$,
the throughput of $\mathrm{UE}_{i,k}$ is given by
\begin{equation}
R_{i,k}=\mathrm{log}\left(1+\frac{\mathrm{SNR}\left|\mathbf{r}_{i,k}^{H}\mathbf{U}_{i}^{H}\mathbf{H}_{i,k}\mathbf{v}_{i,k}\right|^{2}}{1+\mathrm{SNR}\sum_{n=1}^{K}\left|\mathbf{r}_{i,k}^{H}\mathbf{U}_{i}^{H}\mathbf{G}_{i',n}\mathbf{v}_{i',n}\right|^{2}}\right),
\end{equation}
where $\mathrm{SNR}=\frac{P}{N_{0}}$ denotes the transmit signal-to-noise
ratio (SNR) and $\sum_{n=1}^{K}\left|\mathbf{r}_{i,k}^{H}\mathbf{U}_{i}^{H}\mathbf{G}_{i',n}\mathbf{v}_{i',n}\right|^{2}$
is considered as the final residual ICI%
\footnote{It is noted that the actual throughput of IA with limited feedback
can almost always be formulated as (2), regardless of the specific
IA-inspired TB vectors \cite{cho_interference_2012,limited_twocell_mac_2012} before quantization.
In addition, we also keep the notation
of residual ICI to highlight the transmission with limited feedback,
even though we bypass the IA-inspired TB vectors employed in \cite{cho_interference_2012,limited_twocell_mac_2012}. %
}. It is observed from (2) that the residual ICI reduces the system
throughput.

\section{Limited Feedback-Based IA Scheme}

\subsection{Transceiver Design with Limited Feedback}

It is noted that $\mathrm{Col}\left(\mathbf{U}_{i}\right)$ is a $K$-dimension
signal subspace where the desired signals of the $K$ users $\left\{ \mathrm{UE}_{i,k}\right\} _{k=1}^{K}$
are further differentiated by $\left\{ \mathbf{r}_{i,k}\right\} _{k=1}^{K}$.
Because of the imperfect IA with limited feedback, $\mathrm{Col}\left(\mathbf{U}_{i}\right)$
is inevitably contaminated by the residual ICI. For analytical tractability,
we introduce the overall residual ICI within $\mathrm{Col}\left(\mathbf{U}_{i}\right)$
at $\mathrm{B}\mathrm{S}_{i}$ as
\begin{equation}
I_{i}=\sum_{n=1}^{K}\left\Vert \mathbf{U}_{i}^{H}\mathbf{G}_{i',n}\mathbf{v}_{i',n}\right\Vert ^{2}{\geq}\sum_{n=1}^{K}\left|\mathbf{r}_{i,k}^{H}\mathbf{U}_{i}^{H}\mathbf{G}_{i',n}\mathbf{v}_{i',n}\right|^{2},
\end{equation}
which is shown to be an upper bound of the final residual ICI power
after applying $\mathbf{r}_{i,k}^{H}$%
\footnote{The inequality in (3) can be easily proved by using the property$\left\Vert \mathbf{a}\right\Vert \left\Vert \mathbf{b}\right\Vert \geq\left|\mathbf{a}^{H}\mathbf{b}\right|$
with $\left\Vert \mathbf{a}\right\Vert =1$.%
}. It is observed that $I_{i}$ contains the receive filter $\mathbf{U}_{i}$
and the set of the $K$ TB vectors, denoted as $\mathbf{V}_{i'}:=\left\{ \mathbf{v}_{i',n}\right\} _{n=1}^{K}$.
Therefore, both $\mathbf{U}_{i}$ and $\mathbf{V}_{i'}$ can be optimized to
minimize $I_{i}$ for an improved system performance under limited feedback.

Unlike the quantization strategies in \cite{cho_interference_2012,limited_twocell_mac_2012},
which independently choose $\mathbf{v}_{i',n}$ from $\mathrm{UE}_{i,k}$'s
individual codebook to approximate its IA-inspired TB
vector designed with perfect CSI, we aims to jointly choose $\mathbf{V}_{i'}$ from an
new compound codebook to directly minimize the overall residual ICI
$I_{i}$. It is noted that although \cite{cho_interference_2012,limited_twocell_mac_2012}
may find the best quantization for each IA-inspired TB vector,
the selected codewords are not able to directly minimize the residual
ICI by jointly optimizing $\mathbf{U}_{i}$ and $\mathbf{V}_{i'}$. Noting the potential to further improve system performance with optimized
$\mathbf{U}_{i}$ and $\mathbf{V}_{i'}$, we base our IA transceiver design directly on
the overall residual ICI without the reference or constraint of the IA-inspired TB as \cite{cho_interference_2012,limited_twocell_mac_2012},
and our limited feedback-based IA can fully utilize the given codebooks
to achieve the minimal overall residual ICI $I_{i}$.

To start with, we aims to transfer the objective $I_{i}$ into a single
variable function only regarding $\mathbf{V}_{i'}$. This is achieved
by first deriving the structure of the optimal $\mathbf{U}_{i}$
with arbitrary $\mathbf{V}_{i'}$. Let $\mathbf{U}_{i}=[\mathbf{u}_{i,1},...,\mathbf{u}_{i,K}]$,
according to the rotation invariance property of Frobenius norm \cite{matrix_analysis},
we have
\begin{equation}
I_{i}\left(\mathbf{U}_{i},\mathbf{V}_{i'}\right)=\sum_{n=1}^{K}\mathbf{u}_{i,n}^{H}\mathbf{A}_{i'}\left(\mathbf{V}_{i'}\right)\mathbf{u}_{i,n},
\end{equation}
where $\mathbf{A}_{i'}\left(\mathbf{V}_{i'}\right):=\mathbf{\tilde{G}}_{i'}\left(\mathbf{V}_{i'}\right)\mathbf{\tilde{G}}_{i'}^{H}\left(\mathbf{V}_{i'}\right)$ is introduced for simplicity
and $\mathbf{\tilde{G}}_{i'}\left(\mathbf{V}_{i'}\right)=[\mathbf{G}_{i',1}\mathbf{v}_{i',1},...,\mathbf{G}_{i',K}\mathbf{v}_{i',K}]\in\mathbb{C}^{N_{r}\times K}$
is the compound interfering channel at $\mathrm{BS}_{i}$ before
further processing. Let us define the ascendingly ordered eigenvalues
of $\mathbf{A}_{i'}\left(\mathbf{V}_{i'}\right)$ as $\lambda_{1}\left(\mathbf{V}_{i'}\right),...,\lambda_{N_{r}}\left(\mathbf{V}_{i'}\right)$
and the corresponding normalized eigenvectors as $\mathbf{w}_{1}\left(\mathbf{V}_{i'}\right),...,\mathbf{w}_{N_{r}}\left(\mathbf{V}_{i'}\right)$.
For any given $\mathbf{V}_{i'}$, $I_{i}\left(\mathbf{U}_{i},\mathbf{V}_{i'}\right)$
is minimized when $\mathbf{u}_{i,n}\left(\mathbf{V}_{i'}\right)=\mathbf{w}_{n}\left(\mathbf{V}_{i'}\right)$,
$n=1,...,K$ \cite{matrix_analysis}, then the optimal $\mathbf{U}_{i}$
can be defined as a function of $\mathbf{V}_{i'}$ as
\begin{equation}
\mathbf{U}_{i}^{*}\left(\mathbf{V}_{i'}\right)=\left[\mathbf{w}_{1}\left(\mathbf{V}_{i'}\right),...,\mathbf{w}_{K}\left(\mathbf{V}_{i'}\right)\right].
\end{equation}
Applying $\mathbf{U}_{i}^{*}\left(\mathbf{V}_{i'}\right)$, $I_{i}$
can be reformulated as
\begin{equation}
I_{i}\left(\mathbf{U}_{i}^{*}\left(\mathbf{V}_{i'}\right),\mathbf{V}_{i'}\right)=I_{i}\left(\mathbf{V}_{i'}\right)=\rho_{K}\left(\mathbf{A}_{i'}\left(\mathbf{V}_{i'}\right)\right).\label{eq:}
\end{equation}
Comparing (4) and (6), it is observed that the original objective
function $I_{i}\left(\mathbf{U}_{i},\mathbf{V}_{i'}\right)$ has been
transferred to $I_{i}\left(\mathbf{V}_{i'}\right)$ with $\mathbf{U}_{i}^{*}\left(\mathbf{V}_{i'}\right)$,
which is a single-variable function regarding $\mathbf{V}_{i'}$ or
the quantized TB vectors $\left\{ \mathbf{v}_{i',n}\right\} _{n=1}^{K}$.

Focusing directly on the objective function $I_{i}\left(\mathbf{V}_{i'}\right)$, we propose a joint quantization strategy to choose the optimal $\mathbf{V}_{i'}$
from a compound codebook. As a very brief
review and preliminary, it is noted that in \cite{cho_interference_2012,limited_twocell_mac_2012}
$\mathbf{v}_{i',n}$ is individually selected (but not optimized) by $\mathrm{BS}_{i}$
from the randomly generated codebook $\mathcal{C}_{i',n}=\left\{ \mathbf{c}_{i',n,1},\mathbf{c}_{i',n,2},...,\mathbf{c}_{i',n,2^{B}}\right\} $
of $\mathrm{UE}_{i,k}$ with the minimum chordal distance criterion,
where $\mathbf{c}_{i',n,m}\in\mathbb{C}^{N_{t}\times1}$ is the normalized
codeword with the index $m\in\left\{ 1,2,...,2^{B}\right\} $ and
$B$ denotes the number of feedback bits per user. Unlike \cite{cho_interference_2012,limited_twocell_mac_2012},
we enlarge the size of the individual codebook of each
user from $2^{B}$ to $2^{KB}$, and jointly select the TB vectors
$\left\{ \mathbf{v}_{i',n}\right\} _{n=1}^{K}$ to directly minimize
$I_{i}$ with a new strategy. More specifically, we group every $m$-th codeword from each
$\mathcal{C}_{i',n}$, $n=1,2,...,K,$ into a compound codeword as
$\mathbf{C}_{i',\left(m\right)}=\left\{ \mathbf{c}_{i',1,m},...,\mathbf{c}_{i',K,m}\right\}$ \cite{joint_Fang},
then we collect the $2^{KB}$ compound codewords into a new codebook
$\mathcal{C}_{i'}=\left\{ \mathbf{C}_{i',\left(1\right)},\mathbf{C}_{i',\left(2\right)},...,\mathbf{C}_{i',\left(2^{KB}\right)}\right\} $
for the joint quantization
\begin{equation}
\mathbf{V}_{i'}^{*}=\left\{ \mathbf{v}_{i',1}^{*},...,\mathbf{v}_{i',K}^{*}\right\} =\arg\min_{\mathbf{V}_{i'}\in\mathcal{C}_{i'}}\rho_{K}\left(\mathbf{A}_{i'}\left(\mathbf{V}_{i'}\right)\right).\label{eq:-1}
\end{equation}
It is worth pointing out that after the joint quantization, each BS
only needs to exchange and fed back one common index $m_{i'}^{*}$
corresponding with $\mathbf{V}_{i'}^{*}$ to all the served $K$ users
at the cost of $KB$ bits, which means the total feedback per cell
is still the same as \cite{cho_interference_2012,limited_twocell_mac_2012}.
Upon receiving $m_{i'}^{*}$, $\mathrm{UE}_{i',k}$ uses $\mathbf{v}_{i',k}^{*}=\mathbf{c}_{i',k,m_{i'}^{*}}$
as the TB vector and $\mathrm{B}\mathrm{S}_{i}$ uses $\mathbf{U}_{i}^{*}\left(\mathbf{V}_{i'}^{*}\right)$
to establish its cascaded receive filter.

Finally, the complexity of the proposed scheme is briefly discussed.
It is observed (7) mainly includes the construction of $\mathbf{A}_{i'}\left(\mathbf{V}_{i'}\right)$
and its singular value decomposition, which cost $2KN_{r}(3N_{t}-1)$
floating point operations (flops) and $O(N_{r}K^{2})$ flops \cite{matrix_computations},
respectively. Since the searching space has a size of $2^{KB}$, it
is easy to estimate the overall complexity of the proposed scheme
as $O(2^{KB}N_{r}K(3N_{t}+K-1))$. Although our scheme is more complicated
than \cite{cho_interference_2012,limited_twocell_mac_2012}, such
computation overhead is still affordable at the BS.
Moreover, our scheme shows significant throughput gain as compared
to \cite{cho_interference_2012,limited_twocell_mac_2012}, which will
be validated in the following sections.

$Remark$ $1$: Intuitively, \cite{cho_interference_2012,limited_twocell_mac_2012}
can be considered as the traditional IA schemes with limited feedback.
More specifically, the IA-inspired TB vectors are designed before
quantization and the achievability of IA mainly relies
on the qualities of quantization. However, such quantization aims
to approach the IA-inspired TB vectors but not necessarily to the
best achievable throughput performance with the given codebooks. In
contrast to \cite{cho_interference_2012,limited_twocell_mac_2012},
we bypass the IA-inspired transceivers before quantization, and we
straightforwardly approach IA in terms of the minimal residual ICI
by finding the most appropriate codewords within the given codebook.
In this sense, our scheme is specifically designed for limited feedback,
and it is therefore a limited feedback-based IA scheme. In addition,
our scheme is not constrained by the feasibility condition for the
IA-inspired transceivers, and can be implemented with flexible antenna
configurations.

\subsection{Performance Analysis}

In this section, we aims to initially analyze the performance the proposed
scheme. For the comparison purpose, we use \cite{cho_interference_2012}
as a reference and set $N_{t}=N_{r}$ accordingly. Then the throughput
of $\mathrm{UE}_{i,k}$ with perfect feedback is given as
\begin{equation}
R_{i,k}^{PFB}=\log_{2}\left(1+\mathrm{SNR}\left|\mathbf{\tilde{r}}_{i,k}^{H}\mathbf{\tilde{U}}_{i}^{H}\mathbf{H}_{i,k}\mathbf{\tilde{v}}_{i,k}\right|^{2}\right),
\end{equation}
where $\mathbf{\tilde{r}}_{i,k}$, $\mathbf{\tilde{U}}_{i}$ and $\mathbf{\tilde{v}}_{i,k}$
are obtained by using the IA-inspired transceiver design of \cite{cho_interference_2012}
without quantization. Since $R_{i,k}^{PFB}$ assumes the ICI-free
scenario, it can be equivalently considered as the ideal throughput
of the proposed scheme for this comparison. Based on $R_{i,k}^{PFB}$
and $R_{i,k}$ in (6), the throughput loss of $\mathrm{UE}_{i,k}$
with our scheme is defined as
\begin{equation}
\Delta R_{i,k}=R_{i,k}^{PFB}-R_{i,k}.
\end{equation}
The following theorem gives a upper bound of $\mathbb{E}[\Delta R_{i,k}]$
to better understand the gains of the proposed scheme.

\textit{Theorem 1:} When $N_{t}=N_{r}$, the upper bound of the average
throughput loss of $\mathrm{UE}_{i,k}$ with the proposed scheme is
\begin{equation}
\mathbb{E}[\Delta R_{i,k}]\leq\log_{2}\left(1+\mathrm{SNR}\mathbb{E}\left[\min_{m\in\{1,...,2^{KB}\}}\rho_{K}\left(\mathbf{A}_{\left(m\right)}\right)\right]\right),
\end{equation}
where $\left\{ \mathbf{A}_{\left(m\right)},\, m\in\{1,...,2^{KB}\}\right\} $
are some i.i.d. complex central Wishart matrices which follow $\mathcal{CW}_{N_{r}}(K,\mathbf{I}_{N_{r}})$.
\begin{IEEEproof}
We first establish the following inequalities,
\begin{eqnarray}
\mathbb{E}[\Delta R_{i,k}] & \leq & \log_{2}\left(1+\mathbb{E}\left[\mathrm{SNR}\sum_{n=1}^{K}\left|\mathbf{r}_{i,k}^{H}\mathbf{U}_{i}^{H}\mathbf{G}_{i',n}\mathbf{v}_{i',n}\right|^{2}\right]\right)\nonumber \\
 & \leq & \log_{2}\left(1+\frac{\mathbb{E}[I_{i}]}{N_{0}}\right)\label{eq:-2}\\
 & = & \log_{2}\left(1+\mathrm{SNR}\mathbb{E}\left[\min_{m\in\{1,...,2^{KB}\}}\rho_{K}\left(\mathbf{A}_{\left(m\right)}\right)\right]\right),\nonumber
\end{eqnarray}
where the first inequality follows (6) of \cite{cho_interference_2012},
the second inequality follows the inequality in (3), and last equation
follows (6) and (7) by treating $\mathbf{A}_{i'}\left(\mathbf{C}_{i',\left(m\right)}\right)$
as $\mathbf{A}_{\left(m\right)}$. Next, we continue to show the distribution
of $\mathbf{A}_{\left(m\right)}$. Since $\mathbf{c}_{i',n,m}$ is
a normalized vector and the entries of $\mathbf{G}_{i',n}$ are i.i.d.
$\mathcal{CN}(0,1)$, the entries of $\mathbf{G}_{i',n}\mathbf{c}_{i',n,m}$
are i.i.d. $\mathcal{CN}(0,1)$. Then it is easy to show that $\mathbf{A}_{\left(m\right)}$
is a complex central Wishart matrix which follows $\mathcal{CW}_{N_{r}}(K,\,\mathbf{I}_{N_{r}})$
\cite{eigenvalue}. Since the codewords $\mathcal{C}_{i'}$
are i.i.d., it is easy to show that $\left\{ \mathbf{\tilde{G}}_{i'}\left(\mathbf{C}_{i',\left(m\right)}\right),\,\mathbf{C}_{i',\left(m\right)}\in\mathcal{C}_{i'}\right\} $
are also i.i.d., and then the proof is finished.
\end{IEEEproof}
Based on Theorem 1 and (9), a lower bound of the average throughput with the proposed scheme
can be obtained as
\begin{align}
\mathbb{E}[R_{i,k}] & \geq\mathbb{E}[R_{i,k}^{PFB}]\label{eq:-3}\\
 & \,\,\,\,\,\,-\log_{2}\left(1+\mathrm{SNR}\cdot\mathbb{E}\left[\min_{m\in\{1,...,2^{KB}\}}\rho_{K}\left(\mathbf{A}_{\left(m\right)}\right)\right]\right)\nonumber
\end{align}
It is noted that the joint density of the eigenvalues of complex central
Wishart matrix is given by (19) of \cite{eigenvalue}. However, the
distribution of $\min_{m\in\{1,...,2^{KB}\}}\rho_{K}\left(\mathbf{A}_{\left(m\right)}\right)$
is still very complicated. Therefore, we resort to the numerical method
to evaluate the lower bound in (12) for the comparison with \cite{cho_interference_2012},
which can be simulated faster than $\mathbb{E}[R_{i,k}]$.

$Remark$ $2$: For the comparison purpose with \cite{cho_interference_2012},
the above analysis only focus on the configuration $N_{t}=N_{r}$.
Although our scheme is applicable with arbitrary $N_{t}$,
it is often assumed that the users have less antennas than the BS.
When $N_{t}<N_{r}$, it is noted the elements in $\left\{ \mathbf{A}_{\left(m\right)},\, m\in\{1,...,2^{KB}\}\right\} $
are not independent, and the statistical analysis is more involved
and is deferred as future work.

\section{Simulation Results}

We first validate the theoretical results related to Theorem 1 and
compare our scheme with \cite{cho_interference_2012}. The system
configuration are set as $N_{t}=N_{r}=3$, $K=2$, and the number
of feedback bits or $B$ is specified with the simulation curves.
In Fig. 2, the inequality (12) are demonstrated, where the simulation
results are obtained by performing the proposed scheme, and the analytical
results are calculated according to the lower bound in (12). It is
shown that the lower bound of the proposed scheme has already exceeded
the achievable system throughput of \cite{cho_interference_2012}.
Therefore, the throughput gains of the proposed scheme are proved.

Then we compare the proposed scheme with \cite{limited_twocell_mac_2012}
under a variety of system configurations. As shown in Fig. 3, given
the same number of feedback bits $B=4,\,6$, the proposed scheme achieves
significant gain over \cite{limited_twocell_mac_2012} with the system
configuration $N_{t}=2$, $N_{r}=3$, $K=2$. Moreover, it is noted
that due to the IA-feasibility constraint, \cite{limited_twocell_mac_2012}
is not applicable with the system configuration $N_{r}=4$, $K=3$
and $N_{t}\leq\frac{3}{4}K$, while our scheme is still applicable
with arbitrary $N_{t}$, as shown in Fig. 4. It is also observed that
the throughput performance of our scheme increase with $N_{t}$ when
$N_{r}$, $K$ and $B$ are fixed, which shows the benefit of extra
transmit antennas at user.

\begin{figure}
\centering \includegraphics[width=5cm]{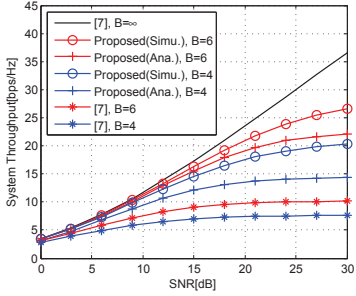} \caption{System throughput comparison with \cite{cho_interference_2012}, $N_{t}=N_{r}=3$,
$K=2$.}
\end{figure}

\begin{figure}
\centering \includegraphics[width=5cm]{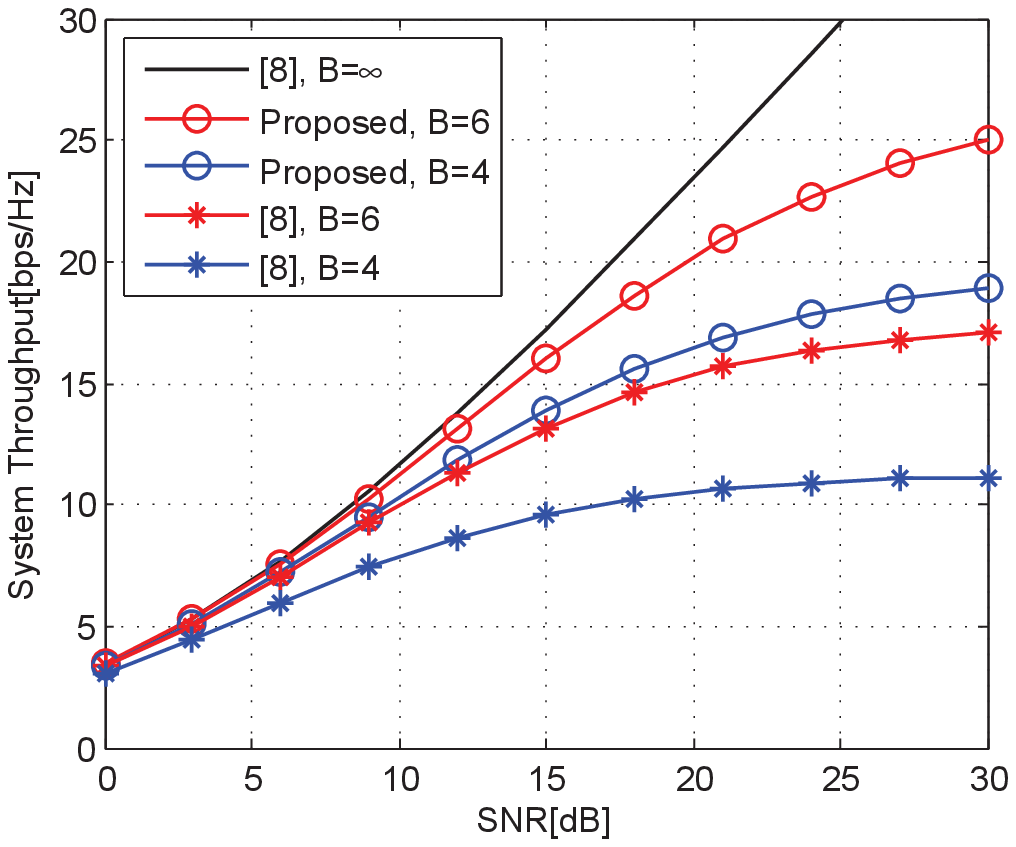} \caption{System throughput comparison with \cite{limited_twocell_mac_2012},
$N_{t}=2$, $N_{r}=3$, $K=2$.}
\end{figure}

\begin{figure}
\centering \includegraphics[width=5cm]{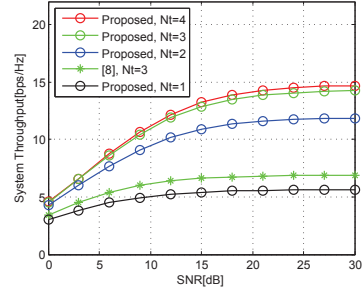} \caption{System throughput comparison with \cite{limited_twocell_mac_2012},
$N_{r}=4$, $K=3$, $B=4$.}
\end{figure}

\section{Conclusion}

In this paper, a limited feedback-based IA scheme has been proposed
for the IMAC. The optimized transceivers have been designed with the
performance-oriented quantization strategy to minimize the residual
ICI. As a beneficial result, the proposed scheme achieves a significant
gain of system throughput and can be implemented with flexible antenna
configurations.

\bibliographystyle{IEEEtran}
\bibliography{biblio}

\end{document}